\documentclass[pre,aps,floatfix,preprint,draft]{revtex4}

\usepackage{color}

\begin{document}

\title{Electro-diffusion in a plasma with two ion species}

\author{Grigory~Kagan\footnote{Email: kagan@lanl.gov} and Xian-Zhu~Tang~\footnote{Email: xtang@lanl.gov}}
\affiliation{Theoretical Division\\
Los Alamos National Laboratory\\
Los Alamos, NM 87545}
\date{\today}

\begin{abstract}
  Electric field is a thermodynamic force that can drive collisional
  inter-ion-species transport in a multicomponent plasma. In an
  inertial confinement fusion (ICF) capsule, such transport causes
  fuel ion separation even with a target initially prepared to have
  equal number densities for the two fuel ion species. Unlike the
  baro-diffusion driven by ion pressure gradient and the
  thermo-diffusion driven by ion and electron temperature gradients,
  electro-diffusion has a critical dependence on the charge-to-mass
  ratio of the ion species. Specifically, it is shown here that
  electro-diffusion vanishes if the ion species have the same
  charge-to-mass ratio. An explicit expression for the electro-diffusion ratio is obtained and used to investigate the relative importance of electro- and baro-diffusion mechanisms. In particular,  it is found
 that electro-diffusion reinforces baro-diffusion in the deuterium and tritium mix, but tends to
  cancel it in the deuterium and helium-3 mix.
\end{abstract}

\maketitle


\section{Introduction}

In inertial confinement fusion (ICF) experiments, where the fuel
assembly is a binary mixture of deuterium and tritium or deuterium and
helium-3, the fusion power production
is proportional to the product
of number densities of the two species.  Optimal fusion yield at a given target temperature in
 local thermodynamic equilibrium requires
not only the fuel assembly be equi-molar for the two ion species, but
also of equal number densities everywhere in the assembly~\cite{note-yield}.  In terms
of the mass densities, this suggests that the mass density of the light
ions
$$
\rho_l = m_l n_l
$$
should be $m_l/m_h$ times of the heavy ion mass density
$$
\rho_h = m_h n_h
$$
for $n_l=n_h.$ Here the subscript ``l'' denotes light ion species and
``h'' denotes heavy ion species.  Defining the mass concentration of the
light ions as
$$
c \equiv \frac{\rho_l}{\rho}
$$
with the mixture ion mass density
$$
\rho \equiv \rho_l + \rho_h,
$$
one finds that the optimal arrangement of equal ion number densities
implies a spatially uniform $c=m_l/(m_l + m_h).$ This condition can
be accurately satisfied in the initial target preparation.  The
dynamical process of implosion, however, can introduce light
and heavy ion separation which degrades the fusion power production~\cite{amendt-prl,amendt-pop, amendt-lapse-rate}.

The collisional inter-ion-species transport or concentration diffusion
is driven by the concentration gradient $\nabla c$ as well as other
thermodynamic forces such as the ion pressure gradient, electron and ion temperature gradients,
and electric field. In the case of a plasma with two species of ions, we show that the diffusive ion mass
flux ${\bf i}$ takes the general form of
$$
{\bf i} = - \rho D \Bigl(\nabla c + k_p\nabla \log p_i + \frac{ek_E}{T_i}\nabla\Phi
+ k_T^{(i)}\nabla\log T_i + k_T^{(e)}\nabla \log T_e\Bigr).
$$
This flux governs $c$ evolution through
\begin{equation}
\label{eq:dcdt}
\rho \frac{\partial c}{\partial t} + \rho {\bf u}\cdot\nabla c + \nabla\cdot{\bf i} = 0,
\end{equation}
where ${\bf u}$ is the plasma fluid velocity.

For the ICF fuel assembly, even when the initial condition has $\nabla
c=0$ by design, the theromodynamic cross terms can drive significant
diffusive flux ${\bf i}$ through the baro-diffusion ($k_p\neq 0$),
electro-diffusion ($k_E\neq 0$), and thermo-diffusion ($k_T^{(e,i)}
\neq 0$).  The above equation attributes baro-diffusion to the total
ion pressure gradient $\nabla p_i,$ which is the sum of the ion
species pressure gradients. This is different from the case of the
neutral gas mixture, where baro-diffusion is considered to be due to
the total mixture pressure. Also, because of the large difference
between the electron and ion masses their temperatures can vary. In
the general case, thermo-diffusion is driven by gradients of both the
electron and ion temperatures.

Baro-diffusion is fundamentally the result of the mass depedence of
the species thermal speed, and we will show it to be indepedent of
electric field and effective gravity due to acceleration and
decceleration in implosions of the ICF target, just like its
counterpart in neutral gas mixtures~\cite{landau-hydro, zeldovich}.
This finding should be contrasted with a previous
result~\cite{amendt-prl, amendt-pop} which suggests gravity and
electric field dependence of $k_p$ for binary plasma mixture. The role
of electric field, or electro-diffusion, is a feature intrinsic to a
plasma.  It is fundamentally the result of the different acceleration
experienced by the ions of different charge-to-mass ratio in an
electric field.  In a low temperature plasma with significant neutral gas
background, this effect is known as ions having
different mobilities~\cite{chen}. We show that
electro-diffusion vanishes ($k_E=0$) if the charge-to-mass ratio is
identical for the two otherwise distinct ion species.  The
charge-to-mass ratio dependence of $k_E$ allows drastically different
electro-diffusion behavior for different binary plasma mixtures. For
the DT mix, we find that
$$
k_E = k_p,
$$
so the electro- and baro-diffusion reinforce each other in a plasma shock.
In contrast, for the D\textsuperscript{3}He mix, one has
$$
k_E = - k_p,
$$
which implies that electro-diffusion tends to cancel baro-diffusion in a plasma shock.

The thermo-diffusion is fundamentally the result of the thermal force in the 
collisional drag between different ion species and between electrons and ions.
The ion-ion thermal force produces a non-vanishing $k_T^{(i)},$ while the ion-electron
thermal force leads to a finite $k_T^{(e)}.$ 
Because of these dependences, thermo-diffusion coefficients require a kinetic calculation of
the said transport coefficients through the distribution function perturbed from a local Maxwellian.
Hence, $k_T^{(i,e)}$ are not thermodynamic quantities like $k_p$ and $k_E$.
It must be noted that with comparable ion masses, the thermal force evaluation is a much more
involved exercise, than the electron-ion one carried out by Braginskii~\cite{braginskii}.
The critical information is nevertheless implicitly contained in standard transport
calculations for multi-component plasmas such as that by Hirshman and Sigmar~\cite{hirshman-sigmar}.
Explicit evaluation of the thermo-diffusion coefficients for the DT and D\textsuperscript{3}He
mixtures will be carried out in a future work. 

The inter-ion-species diffusion, which modifies the relative number
density of the two fusion reactants, has been an issue of interest in
inertial confinement fusion and dense plasma research.  Among those we
are familiar with, C.~H.~Chang, B.~Albright, and W.~Daughton from Los
Alamos National Laboratory have investigated the models for ${\bf i}$
with varying degrees of approximation for evaluating the baro- and
thermo-diffusion in the past decade, albeit in unpublished reports.
More recently, it has been attracting special attention for the
pioneering analysis of Amendt et al~\cite{amendt-prl,amendt-pop},
which shows baro-diffusion may be responsible for the discrepancy
between the neutron yield measured during ICF implosions and that
predicted by simulations.  This can largely be attributed to the
realization that the strong pressure and temperature gradients, along
with the strong electric field~\cite{rygg-science}, are induced by
shock waves inevitably present in the imploded capsule.
Inter-diffusion between the two ion species must therefore take place;
the resulting separation of the fuel constituents in the hot spot can
significantly degrade the fusion
yield~\cite{amendt-prl,amendt-pop}. An experimental evidence for the
fuel stratification in ICF implosions has been reported by Casey et al
based on their measurements at the OMEGA laser
facility~\cite{casey-prl}.

The purpose of this paper is to clarify the underlying diffusion
mechanisms, especially the role of electro-diffusion in relation to
baro-diffusion. The derivation of ${\bf i}$ based on multi-component
collisional fluid models also provides a framework for incorporating
this important physics in ICF modeling.  As highlighted earlier for
a binary plasma mix, there are a number of subtleties absent in neutral
gas mixture, and not addressed in preceding studies.  Since the
inter-species diffusion is a general topic for multi-component
plasmas, it is expected that the results described here would be of
general interest, for example, to tokamak edge plasma modeling, plasma processing,
and stellar structures. 

The rest of the paper is organized as follows. In the next section,
the thermodynamic framework for evaluating electro-diffusion is
outlined. In section III, the momentum equation for an ion species in the
center-of-mass frame is obtained and simplified by imposing an
ordering relevant to a collisional plasma shock. Then, in section IV,
this equation is utilized to write the diffusive mass flux of an ion
species in the general form of section II, thereby explicitly
evaluating the electro-diffusion ratio. Finally, in section V,
implications of electro-diffusion in the ICF context are discussed.

\section{Thermodynamic expression for the diffusive flux}

To define the diffusive mass flux of ion component $\alpha$, the
center-of-mass velocity ${\bf u}$ is first introduced by
\begin{equation}
\label{eq: c-m-velocity}
\rho {\bf u} \equiv \sum_{\alpha} \rho_{\alpha} {\bf u}_{\alpha},
\end{equation}
where $\rho_{\alpha}$ is the partial density of component $\alpha$ and
$\rho= \sum_{\alpha}\rho_{\alpha} $ is the mixture total ion
density. Also, ${\bf u}_{\alpha}$ denotes the net flow velocity of the
component and the sum on the right side of Eq.~(\ref{eq:
  c-m-velocity}) is over all the components present in the mix. The
diffusive mass flux is then given by
\begin{equation}
\label{eq: flux-def}
{\bf i}_{\alpha}= \rho_{\alpha}( {\bf u}_{\alpha}-{\bf u}).
\end{equation}

If the system is close to local thermodynamic equilibrium, a linear
relation between the thermodynamic forces and the resulting fluxes can
be assumed. For a neutral gas mixture, the total diffusive mass flux
of component $\alpha$ can be written as~\cite{landau-hydro,zeldovich}
\begin{equation}
\label{eq: canonical-flux}
{\bf i}_{\alpha} = - \rho D \Bigl( \nabla c_{\alpha} +k_p \nabla \log{p}  + k_T \nabla \log{T} \Bigr),
\end{equation}
where $p$ and $T$ denote the mixture total pressure and temperature, respectively, and $c_{\alpha} \equiv \rho_{\alpha}/\rho$ denotes concentration of the component $\alpha$. Parameter $D$ is called diffusion coefficient; it governs the diffusive flux when only the concentration gradient is present. In view of Eq.~(\ref{eq: canonical-flux}), baro- and thermo-diffusion coefficients are then equal to $k_p D$ and $k_T D$, respectively. Dimensionless parameters $k_p$ and $k_T$ are usually referred to as baro- and thermo-diffusion ratios, respectively.

Interestingly, $k_p$ is a thermodynamic quantity, i.e. it can be
evaluated given local values of thermodynamic variables and does not
depend on the details of collisions~\cite{landau-hydro}. In
particular, for a binary mix it can be found~\cite{landau-hydro,
  zeldovich}
\begin{equation}
\label{eq: baro-diff-ratio}
k_p = c(1-c)(m_h-m_l)\Bigl( \frac{c}{m_l} +  \frac{1-c}{m_h}    \Bigr),
\end{equation}
where $m_l$ and $m_h$ are the atomic masses of the light and heavy
fractions, respectively. Also, $c \equiv c_l$ stands for the
concentration of the light fraction; the subscript "l" is dropped to
simplify notation, as in a binary mix the concentration of the heavy
fraction can be recovered through $c_h=1-c_l$. In contrast to $k_p,$
$k_T$ is an intrinsically kinetic quantity and is subject to change
depending on the collisional model.

To account for the effect of the electric field, the corresponding
force needs to be added on the right side of Eq.~(\ref{eq:
  canonical-flux}) to rewrite it as
\begin{equation}
  \label{eq: canonical-flux-1}
  {\bf i}_{\alpha} = - \rho D \Bigl( \nabla c_{\alpha} +k_p \nabla \log{p}  + k_T \nabla \log{T} + \frac{e k_E}{T}\nabla \Phi \Bigr),
\end{equation}
where $\Phi$ is the electrostatic potential and, by analogy with $k_p$
and $k_T$, the electro-diffusion ratio $k_E$ is introduced. In what
follows, we focus on the case of a binary plasma mix, i.e. plasma
consisting of two ion species and electrons. To illuminate the new
features brought by the electric field, as compared to the case of a
binary mix of neutral gases, we apply Eq.~(\ref{eq: canonical-flux-1})
to the system consisting of the two ion species. Within such an
approach, the electron species is viewed as an external factor that
affects the system of interest through the electric field and
collisions. In other words, the electrons do not make contribution in
the definition of $\rho, {\bf u},$ and $p,$ that is consistent with
our objective to understand the relative motion of the two ion
species. It is worth noticing that because of small inertia the electron
contribution to the overall plasma density and flow is
negligible. Thus, for all practical purposes $\rho$ and ${\bf u}$ can
still be referred to as the plasma density and flow, respectively. On
the contrary, the electron and ion pressures are generally comparable
and employing the total ion pressure, rather than the overall plasma
pressure, in place of $p$ in Eq.~(\ref{eq: canonical-flux-1}) is
crucial.

Assuming the diffusive flux of the form~(\ref{eq: canonical-flux-1}) it is possible to evaluate $k_E$ by generalizing formal thermodynamic 
methods used in Ref.~\cite{landau-hydro} to evaluate $k_p$~\cite{amendt-lapse-rate}. 
Instead, here we start from the first-principle based momentum
conservation equations for individual species to automatically recover this form. Importantly, in addition to readily providing $k_p$ and $k_E$, this technique gives formulae for $D$, $k_T^{(i)}$ and $k_T^{(e)}$ in terms of standard transport coefficients. In so doing, it lays the framework for evaluating the overall effect of the ion concentration diffusion that is inherently not possible within the thermodynamic approach.

\section{Momentum conservation for ion species}

We start by writing the momentum equations for the two ion species
\begin{equation}
\label{eq: momentum}
\rho_{\alpha}\frac{d_{\alpha} {\bf u}_{\alpha}}{dt} +\nabla \cdot \tensor{P}_{\alpha}-n_{\alpha} Z_{\alpha}e{\bf E} - \rho_{\alpha}{\bf F}_{\alpha}  = \sum_{\beta}{\bf R}_{\alpha\beta}  ,
\end{equation}
where the subscript $\alpha$ can be "l" and "h" to denote the light
and heavy ion species, respectively. In Eq.~(\ref{eq: momentum})
$n_{\alpha}$, $ p_{\alpha}$ and $Z_{\alpha}$ stand for the species'
number density, partial pressure and charge number,
respectively. Also, ${\bf E}$ stands for the electric
field. Acceleration due to an external force of a non-electric origin,
such as the gravitational force, is denoted by ${\bf F}_{\alpha}$,
while ${\bf R}_{\alpha\beta}$ is the force density due to collisional
momentum exchange with the species $\beta$ and the sum on the right
side of Eq.~(\ref{eq: momentum}) is over all plasma species, including electrons. The
pressure tensor $\tensor{P_{\alpha}}$ is defined in the frame
co-moving with the species net flow by 
$$
\tensor{P}_{\alpha} \equiv
m_{\alpha} \int ({\bf v}-{\bf u}_{\alpha}) ({\bf v}-{\bf
  u}_{\alpha})f_{\alpha} d^3v,
$$ 
where $f_\alpha$ is the species
distribution function. Finally,
$$
d_{\alpha}/dt
\equiv \partial/\partial t +{\bf u}_{\alpha}\cdot\nabla. 
$$

Next, we obtain the momentum equation for the center of mass velocity
defined by Eq.~(\ref{eq: c-m-velocity}). To do so, it is convenient to
rewrite Eq.~(\ref{eq: momentum}) in the conservative form:
\begin{equation}
\label{eq: momentum-conservative}
\frac{\partial (\rho_{\alpha} {\bf u}_{\alpha})}{\partial t} + \nabla\cdot (\rho_{\alpha} {\bf u}_{\alpha}{\bf u}_{\alpha}) +  \nabla \cdot \tensor{P}_{\alpha}-n_{\alpha} Z_{\alpha}e{\bf E} - \rho_{\alpha}{\bf F}_{\alpha}  = \sum_{\beta}{\bf R}_{\alpha\beta},
\end{equation}
where species continuity equation 
$$
d_{\alpha}\rho_{\alpha}/dt + \rho_{\alpha}
\nabla \cdot {\bf u}_{\alpha}=0
$$ 
is used. Introducing the species'
velocity in the center-of-mass frame 
\begin{equation}
\label{eq:w-def}
{\bf w}_{\alpha} \equiv {\bf
  u}_{\alpha} - {\bf u},
\end{equation}
noticing that 
$$ 
\nabla\cdot (\rho_{\alpha}
{\bf u}_{\alpha}{\bf u}_{\alpha}) = \nabla\cdot (\rho_{\alpha} {\bf
  u}{\bf u}) + \nabla\cdot (\rho_{\alpha} {\bf w}_{\alpha}{\bf u}) +
\nabla\cdot (\rho_{\alpha} {\bf u}{\bf w}_{\alpha}) + \nabla\cdot
(\rho_{\alpha} {\bf w}_{\alpha}{\bf w}_{\alpha})
$$ 
and summing
Eq.~(\ref{eq: momentum-conservative}) over all the ion species we find
\begin{equation}
\label{eq: momentum-c.m.-conservative}
\frac{\partial (\rho{\bf u})}{\partial t} + \nabla\cdot (\rho{\bf u}{\bf u})  +\nabla \cdot \tensor{P}_i- \sum_{\alpha = i_j}(n_{\alpha} Z_{\alpha}e{\bf E} + \rho_{\alpha}{\bf F}_{\alpha})  = \sum_{\alpha = i_j}{\bf R}_{\alpha e} ,
\end{equation}
where we use that $ \sum_{\alpha = i_j}\rho_{\alpha} {\bf w}_{\alpha}
= 0$, $\tensor{P}_i = \sum_{\alpha}(\tensor{P}_{\alpha} +
\rho_{\alpha} {\bf w}_{\alpha}{\bf w}_{\alpha})$ is the total ion
pressure tensor in the center-of-mass frame and subscript "$\alpha =
i_j$" denotes summation over the ion species only.  Equation~(\ref{eq:
  momentum-c.m.-conservative}) is then easy to transform to a more
familiar form
\begin{equation}
\label{eq: momentum-c.m.}
\rho\frac{d{\bf u}}{dt} + \nabla p_i + \nabla \cdot \tensor{\Pi}_i- \sum_{\alpha = i_j}(n_{\alpha} Z_{\alpha}e{\bf E} + \rho_{\alpha}{\bf F}_{\alpha} + {\bf R}_{\alpha e})  = 0,
\end{equation}
where $d/dt \equiv \partial/\partial t +{\bf u}\cdot\nabla $ and $p_i$
and $\tensor{\Pi}_i$ are the total ion pressure and viscous stress
tensor, respectively.

Finally, we rewrite equations for the individual species flows in the center-of-mass frame to obtain
\begin{equation}
\label{eq: momentum-1}
\frac{d\rho_{\alpha}{\bf w}_{\alpha}}{dt} +\rho_{\alpha} {\bf w}_{\alpha} \cdot \nabla {\bf u} +\rho_{\alpha} {\bf w}_{\alpha} \nabla \cdot {\bf u}+
\nabla \cdot \tensor{P}_{\alpha}^{c.m}-n_{\alpha} Z_{\alpha}e{\bf E} - \rho_{\alpha}{\bf F}_{\alpha}  + \rho_{\alpha}\frac{d{\bf u}}{dt}=  \sum_{\beta}{\bf R}_{\alpha\beta} ,
\end{equation}
where 
$$
\tensor{P}_{\alpha}^{c.m} = \tensor{P}_{\alpha} + \rho_{\alpha}
{\bf w}_{\alpha}{\bf w}_{\alpha} =m_{\alpha} \int ({\bf v}-{\bf u}) ({\bf
  v}-{\bf u})f_{\alpha} d^3v
$$ 
is the species' pressure tensor in the
center-of-mass frame. By splitting a scalar pressure out
of $\tensor{P}_{\alpha}^{c.m}$, Eq.~(\ref{eq: momentum-1}) is then
rewritten further to find
\begin{equation}
\label{eq: momentum-2}
\frac{d\rho_{\alpha}{\bf w}_{\alpha}}{dt} +\rho_{\alpha} {\bf w}_{\alpha} \cdot \nabla {\bf u} +\rho_{\alpha} {\bf w}_{\alpha} \nabla \cdot {\bf u}+\nabla p_{\alpha}^{c.m}+
\nabla \cdot \tensor{\Pi}_{\alpha}^{c.m}-n_{\alpha} Z_{\alpha}e{\bf E} - \rho_{\alpha}{\bf F}_{\alpha}  + \rho_{\alpha}\frac{d{\bf u}}{dt}= \sum_{\beta}{\bf R}_{\alpha\beta},
\end{equation}
where $p_{\alpha}^{c.m}$ and $\tensor{\Pi}_{\alpha}^{c.m}$ are the
species' partial pressure and viscous stress tensor, respectively,
evaluated in the center-of-mass frame.

The calculation presented here is motivated by the problem of the ion
species diffusion within an ICF relevant shock wave front. In general,
the shock front width can be as small as the mean free path, making a
local treatment, as well as the framework of the previous section,
invalid. However, for a moderately strong shock, the front width can
be assumed much greater than the mean free path $\lambda$, i.e.
\begin{equation}
\label{eq: ordering-space}
\lambda/\Delta << 1,
\end{equation}
where $\Delta$ is the characteristic spatial scale of the plasma (e.g. the shock width). The mean free paths can substantially differ for pre- and post-shocked plasmas; for definitiveness, we refer $\lambda$ to the post-shock mean free path. Ordering~(\ref{eq: ordering-space}) is usually satisfied for Mach numbers $\lesssim 2$ and ensures that the plasma remains mostly collisional throughout the shock front~\cite{jaffrin}.  
 The
characteristic temporal scale $\tau$ can then be estimated from
$\tau^{-1} \sim v_{sh}/\Delta$, where $v_{sh}$ is the shock
speed. Assuming that ion masses are comparable across
different species and introducing the characteristic ion thermal speed
$v_{th-i} \sim v_{sh}$ we find
\begin{equation}
\label{eq: ordering-time}
\nu_i \tau \sim (\lambda/\Delta)^{-1}>> 1,
\end{equation}
 where $\nu_{i} \sim  v_{th-i}/\lambda$ is the characteristic ion collision frequency.
 
Estimating the friction between the ion species by $
\mu_{lh} \nu_{i} n_l (w_l - w_h)$, where $\mu_{lh}$ is the reduced
mass for the light and heavy ions, it is straightforward to show
 that ordering~(\ref{eq: ordering-time}) makes 
$$
\frac{w_{\alpha}}{v_{th-i}} \sim \frac{\lambda}{\Delta} \ll 1,
$$ 
thereby ensuring that the system is close
 to a local equilibrium. 
The terms on the left side of Eq.~(\ref{eq: momentum-2}) that contain
both the spatial gradient and $w_{\alpha}$ are quadratic in the small
parameter $\lambda/\Delta$ and can be dropped. For the same reason
$\tensor{P}_{\alpha}^{c.m} \approx \tensor{P}_{\alpha}$ and the
superscript "$c.m.$" appearing next to the partial pressure and
viscous tensor can be omitted. Finally, due to the same estimate for the friction, the $\partial {\bf
  w}_{\alpha}/\partial t$ term can be dropped as well and
Eq.~(\ref{eq: momentum-2}) reduces to
\begin{equation}
\label{eq: quasi-equlibrium}
\nabla p_{\alpha}+\nabla \cdot \tensor{\Pi}_{\alpha}-n_{\alpha} Z_{\alpha}e{\bf E} - \rho_{\alpha}{\bf F}_{\alpha}  + \rho_{\alpha}\frac{d{\bf u}}{dt}=  \sum_{\beta}{\bf R}_{\alpha\beta}.
\end{equation}
Equation~(\ref{eq: quasi-equlibrium}) is valid for a plasma with an
arbitrary number of species as long as ordering~(\ref{eq: ordering-time}) is
obeyed. In the next section, we apply it to evaluate the
electro-diffusion coefficient in a plasma with two ion species.

\section{Evaluating electro-diffusion coefficient}

The viscous term appearing on the left side of Eq.~(\ref{eq:
  quasi-equlibrium}) is governed by the second order derivatives of
macroscopic parameters. Its contribution is therefore not retained in
Eq.~(\ref{eq: canonical-flux-1}), which is obtained by assuming linear
relation between the thermodynamic forces and the resulting flux.  In
principle, this contribution may be substantial and effectively modify
the baro-diffusion ratio~\cite{landau-hydro, zeldovich}. However, as
the main goal of the present study is to elucidate the role of the
electric field on the ion diffusion, in what follows we drop $ \nabla
\cdot \tensor{\Pi}_{\alpha}$ on the left side of Eq.~(\ref{eq:
  quasi-equlibrium}).  Then, employing Eq.~(\ref{eq: momentum-c.m.}) with the $ \nabla
\cdot \tensor{\Pi}_{i}$ term also dropped
to evaluate $du/dt$ in Eq.~(\ref{eq: quasi-equlibrium}) we find
\begin{eqnarray}
 \nonumber
 \sum_{\beta=i_j}{\bf R}_{\alpha\beta} + ( {\bf R}_{\alpha e}-\frac{\rho_{\alpha}}{\rho} \sum_{\beta=i_j} {\bf R}_{\beta e})=\\
\label{eq: diffusion}
( \nabla p_{\alpha}-\frac{\rho_{\alpha}}{\rho}\nabla p_i)  - (Z_{\alpha} n_{\alpha}  -\frac{\rho_{\alpha}}{\rho} \sum_{\beta=i_j} Z_{\beta} n_{\beta})e {\bf E} - ( \rho_{\alpha} {\bf F}_{\alpha}-\frac{\rho_{\alpha}}{\rho} \sum_{\beta=i_j} \rho_{\beta} {\bf F}_{\beta}).
\end{eqnarray}
Notice, that if we had included electrons into the system, the sum
over $\beta$ in the second term on the right side of Eq.~(\ref{eq:
  diffusion}) would vanish due to quasi-neutrality.

Equation~(\ref{eq: diffusion}) gives the light ion species diffusion
velocity ${\bf w}_l$ through the ${\bf R}_{lh}$ dependence on the net
velocity difference between the ion species since
\begin{equation}
\label{eq: diffusion-velocity}
{\bf w}_l-{\bf w}_h = \Bigl( 1+\frac{\rho_l}{\rho_h} \Bigr){\bf w}_l = \frac{{\bf w}_l}{1-c} ,
\end{equation}
where $c$ is the concentration of the light ion species. In a
multi-component plasma~\cite{hirshman-sigmar}, the collisional drag
between species $\alpha$ and $\beta$
\begin{equation}
\label{eq: friction-binary}
{\bf R}_{\alpha\beta} =-[A_{\alpha\beta}\mu_{\alpha\beta} n_{\alpha} \nu_{\alpha\beta} ({\bf w}_{\alpha} - {\bf w}_{\beta})+c^{(1)}_{\alpha\beta} n_{\alpha}\nabla T_{\alpha} +c^{(2)}_{\alpha\beta} n_{\beta}\nabla T_{\beta}] ,
\end{equation}
where, in general, coefficients $A_{\alpha\beta}$,
$c^{(1)}_{\alpha\beta}$ and $c^{(2)}_{\alpha\beta}$ are complicated
functions of the masses, densities and charge numbers of all the
species and ${\bf R}_{\alpha\alpha}=0$ for arbitrary $T_\alpha$
implies $c^{(1)}_{\alpha\alpha} = c^{(2)}_{\alpha\alpha}=0$. Also,
$T_{\alpha}$ is the temperature of species $\alpha$,
$\mu_{\alpha\beta}\equiv m_{\alpha}m_{\beta}/(m_{\alpha}+m_{\beta})$
is the reduced mass and $\nu_{\alpha\beta}$ stands for the frequency
of collisions between species ${\alpha}$ and ${\beta}$.
Conventionally, the terms on the right side of Eq.~(\ref{eq:
  friction-binary}) proportional to the velocity difference and
temperature gradients are referred to as the frictional and thermal
forces, respectively. Summing Eq.~(\ref{eq: friction-binary}) over the
ion species we find
\begin{equation}
\label{eq: friction}
\sum_{\beta=i_j}{\bf R}_{\alpha\beta} =-\sum_{\beta=i_j} [A_{\alpha\beta}\mu_{\alpha\beta} n_{\alpha} \nu_{\alpha\beta} ({\bf w}_{\alpha} - {\bf w}_{\beta})+B_{\alpha\beta} n_{\beta}\nabla T_{\beta}] ,
\end{equation}
where $B_{\alpha\alpha} \equiv \sum_{\beta=i_j} c^{(1)}_{\alpha\beta}$ and $B_{\alpha\beta} \equiv c^{(2)}_{\alpha\beta}$ for $\alpha \neq \beta$.

When the elementary masses of species $\alpha$ and $\beta$ are
comparable, the thermal force acting between them depends on both
$\nabla T_{\alpha}$ and $\nabla T_{\beta},$ i.e. $c^{(1)}_{\alpha\beta} \sim c^{(2)}_{\alpha\beta}$ on the right side of Eq.~(\ref{eq:
  friction-binary}). In contrast, the thermal force acting between the
electron and any of the ion species is dominated by the electron
temperature gradient, because the thermal speed of electrons is much
greater than that of ions, {\it i.e.},
\begin{equation}
\label{eq:i-e-friction-1}
{\bf R}_{\alpha e} = - [ A_{\alpha e} \mu_{\alpha e} n_\alpha \nu_{\alpha e} ({\bf w}_\alpha - {\bf w}_e)
+ B_{\alpha e} n_e \nabla T_e]
\end{equation}
where we set $B_{\alpha e} \equiv c^{(2)}_{\alpha e}$ to unify notation with Eq.~(\ref{eq: friction}), and
$$
{\bf w}_e \equiv {\bf u}_e - {\bf u}
$$
is the electron flow velocity in the ion center-of-mass flow frame.
Unlike ${\bf w}_{l,h} \sim (\lambda/\Delta) v_{th-i}$ from our
collisional ordering, there is no such constraint on the electron flow
${\bf w}_e$ due to the much larger electron thermal velocity.  This is consistent
with the well-known result that a collisional plasma can carry a
substantial current in the electron channel despite that the ion current is
negligibly small in the short mean-free-path limit.  To estimate the
ion-electron frictional force on the right side of
Eq.~(\ref{eq:i-e-friction-1}) we introduce the plasma current
\begin{equation}
\label{eq: current-def}
{\bf J} \equiv -en_e{\bf u}_e + \sum_{\beta=i_j} en_\alpha Z_\alpha {\bf u}_\alpha =
 -en_e{\bf w}_e + \sum_{\beta=i_j} en_\alpha Z_\alpha {\bf w}_\alpha,
\end{equation}
where the quasi-neutrality condition along with Eq.~(\ref{eq:w-def}) is used to obtain the right side of the equation. Then, using 
Eq.~(\ref{eq: diffusion-velocity}), the friction between the light ions and electrons can be rewritten as
\begin{equation}
\label{eq:i-e-friction-2}
{\bf R}_{l e}^f = - A_{l e} \mu_{l e} n_l \nu_{l e} ({\bf w}_l - {\bf w}_e) = 
- A_{l e} \mu_{l e} n_l \nu_{l e} \Bigl[ \frac{Z_h/m_h}{cZ_l/m_l + (1-c)Z_h/m_h}{\bf w}_l  +\frac{1}{en_e} {\bf J} \Bigr].
\end{equation}
The term on the right side involving ${\bf w}_l $ is smaller than the
friction between the ion species by a factor of
$\sqrt{m_e/m_{l,h}}$. In an ambipolar plasma ${\bf J}=0$ and the
frictional force can be neglected on the right side of
Eq.~(\ref{eq:i-e-friction-1}). Moreover, even for a plasma carrying
significant current through the electrons due to the $-en_e{\bf w}_e$ term in Eq.~(\ref{eq: current-def}),
 the ion-electron friction
force is much less than its ion-ion counterpart as long as
\begin{equation}
\label{eq:current-constraint}
\frac{J}{en_e v_{th-i}} \ll \frac{\lambda}{\Delta}\sqrt{\frac{m_{l,h}}{m_e}}.
\end{equation}

Condition~(\ref{eq:current-constraint}) is the most restrictive in the
case of a weak shock, where the shock front width can be many times of the
ion-ion mean free path making $(\lambda/\Delta)\sqrt{m_{l,h}/m_e}$ of
order unity~\cite{jaffrin}. The constraint on the plasma current
becomes $J \ll en_e v_{th-i}.$ Hence, in the absence of large currents
on the order of $e n_e v_{th-i}$ or greater, ${\bf R}_{\alpha e}^f$
can be ignored for an ICF relevant shock wave. Consequently, the
ion-electron collisional drag is dominated by the thermal force,
\begin{equation}
\label{eq: el-thermal-force}
{\bf R}_{\alpha e}\approx -B_{\alpha e}n_e\nabla T_e.
\end{equation}

Applying general expressions~(\ref{eq: friction})~and~(\ref{eq:
  el-thermal-force}) to our case and setting $\alpha=l$ the left side
of Eq.~(\ref{eq: diffusion}) is now evaluated to find
\begin{eqnarray}
 \nonumber
 \sum_{\beta=i_j}{\bf R}_{l\beta} + ( {\bf R}_{le}-\frac{\rho_{l}}{\rho} \sum_{\beta=i_j} {\bf R}_{\beta e}) = \\
 \label{eq: friction-1}
 -A_{lh}\mu_{lh} n_l \nu_{lh} ({\bf w}_l - {\bf w}_h)    -    \sum_{\beta=l,h} B_{l\beta} n_{\beta}\nabla T_{\beta} 
 - (B_{le} - c \sum_{\beta=l,h}B_{\beta e})n_e\nabla T_e.
\end{eqnarray}
While the coefficients $B_{\beta e}$ are relatively easy to recover by
generalizing the corresponding Braginskii's result for a simple
plasma~\cite{braginskii}, evaluating $A_{lh}$, $B_{lh}$ and $B_{ll}$
is quite complicated even in the case of only two different ion
species with comparable masses, charge numbers and
concentrations. Fortunately, it will be found unnecessary for the
purpose of this paper, so we proceed leaving coefficients of
Eq.~(\ref{eq: friction-1}) unspecified.

To complete the calculation, Eq.~(\ref{eq: diffusion}) needs to be
rewritten in the canonical form~(\ref{eq: canonical-flux-1}). The
terms on the right side of Eq.~(\ref{eq: diffusion}) then have to be
expressed in terms of the total ion pressure $p_i$ and the light
species concentration $c$. We now proceed by doing so in the first
term on the right side of Eq.~(\ref{eq: diffusion}) that is
responsible for baro-diffusion.

First, we observe that energy exchange \emph{between} the ion species
with comparable masses takes place over the same time scale as thermal
equilibration \emph{within} any of the two species. Hence, under
ordering~(\ref{eq: ordering-time}), $T_l\approx T_h$ and the overall
ion temperature $T_i$ can be introduced. Next, we notice that
$p_i=(n_l+n_h)T_i=(\rho_l/m_l+\rho_h/m_h)T_i$, where $m_l$ and $m_h$
are the light and heavy ion masses, respectively, to obtain
\begin{equation}
\label{eq: partial-pressure}
p_l=\frac{cm_h}{cm_h+(1-c)m_l} p_i,
\end{equation}
and
\begin{equation}
\label{eq: partial-pressure-grad}
\nabla p_l=\frac{cm_h}{cm_h+(1-c)m_l}\nabla p_i+\frac{p_im_hm_l}{[cm_h+(1-c)m_l]^2} \nabla c.
\end{equation}
The first term on the right side of Eq.~(\ref{eq: diffusion}) is then evaluated to find
\begin{equation}
\label{eq: baro-diffusion}
( \nabla p_l-\frac{\rho_l}{\rho}\nabla p_i)=\frac{\rho T_i}{cm_h+(1-c)m_l}\Bigl[\nabla c +c(1-c)(m_h-m_l)\Bigl(\frac{c}{m_l}+\frac{1-c}{m_h}\Bigr)\nabla \log{p_i}\Bigr].
\end{equation}

Expression inside the square brackets of Eq.~(\ref{eq:
  baro-diffusion}) is normalized, i.e. the coefficient in front of the
$\nabla c$ term is equal to unity. In view of Eq.~(\ref{eq:
  canonical-flux-1}) it means that the coefficient in front
of the $\nabla \log{p_i}$ term is equal to the baro-diffusion ratio
$k_p$.  Importantly, this ratio, obtained here from ion fluid
equations, matches the result~(\ref{eq: baro-diff-ratio}), found in
Refs.~\cite{landau-hydro, zeldovich} for a binary mix of ideal
gases. Of course, this is just a reflection of the aforementioned fact
that $k_p$ is a thermodynamic quantity and does not depend on the
details of the collisional exchange between the species. It should be
noted that recovering the same $k_p$ as in Refs.~\cite{landau-hydro,
  zeldovich} manifests the key difference between our approach and
that of Refs.~\cite{amendt-prl, amendt-pop}, where $k_p$ is found to be
dependent upon the electric field.

With the technique presented in the preceding paragraphs, $k_E$ can be
straightforwardly calculated in the same way as $k_p$. Before doing
so, we take a brief detour and apply this technique to clarify the
role of gravity. The effective gravity appears in ICF relevant
problems when acceleration of the capsule during implosion needs to be
accounted for. Upon switching to the frame co-moving with the capsule
the inertial force enters the momentum equation that is formally
equivalent to placing the system into external field with
an effective gravitational acceleration ${\bf g}$.

In Refs.~\cite{amendt-prl, amendt-pop}, gravity is found to modify the
expression for $k_p$, so does the electric field. Within the
framework of the present study, gravity can be included by setting the
external force ${\bf F}_\alpha$ equal to ${\bf g}$~~for both
$\alpha=l$ and $\alpha=h$. The third term on the right side of
Eq.~(\ref{eq: diffusion}) is then found to vanish; that is, the
gravitational force does not drive a diffusive flux. This
result obtained with a rather formal method has a trivial physical
explanation. Namely, gravity gives the same acceleration to all ions
regardless of their mass and charge number and therefore introducing
it into otherwise unchanged system does not directly contribute to the
species separation. Of course, gravity can still affect ion
concentrations indirectly. For example, it can do so by modifying the
electron pressure balance. The electric field then has to adjust,
thereby modifying the ion flux through its electro-diffusive
component.

Now we obtain the electro-diffusion ratio by writing the total
diffusive flux of the light ion species in the canonical form. The
second term on the right side of Eq.~(\ref{eq: diffusion}) is first
evaluated to find
\begin{equation}
\label{eq: electro-diffusion}
(Z_l n_l  -\frac{\rho_l}{\rho} \sum_{\beta=i_j} Z_{\beta} n_{\beta})e {\bf E} = \rho c(1-c) \Bigl( \frac{Z_l}{m_l} - \frac{Z_h}{m_h}  \Bigr)e {\bf E}.
\end{equation}
Next, the terms on the right side of Eq.~(\ref{eq: diffusion}) are
collected with the help of Eqs.~(\ref{eq:
  baro-diffusion})~and~(\ref{eq: electro-diffusion}) and Eq.~(\ref{eq:
  friction-1}) is employed along with Eq.~(\ref{eq:
  diffusion-velocity}) to find
\begin{equation}
\label{eq: canonical-2}
{\bf i}_l \equiv \rho_l {\bf w}_l =
- \rho D \Bigl( \nabla c +k_p \nabla \log{p_i} + \frac{e k_E}{T_i}\nabla \Phi + k_T^{(i)} \nabla \log{T_i}  + k_T^{(e)} \nabla \log{T_e}\Bigr),
\end{equation}
where, as recovered by Eq.~(\ref{eq: baro-diffusion}), $k_p$ is still given by Eq.~(\ref{eq: baro-diff-ratio})  and
\begin{eqnarray}
\label{eq: diffusion-coeff}
D= \frac{\rho T_i}{A_{lh}\mu_{lh} n_l \nu_{lh}} \times \frac{c(1-c)}{cm_h+(1-c)m_l}, \\
\label{eq: electro-diff-ratio}
k_E = m_lm_h c(1-c)  \Bigl( \frac{c}{m_l} +  \frac{1-c}{m_h}    \Bigr)  \Bigl( \frac{Z_l}{m_l} - \frac{Z_h}{m_h}  \Bigr),\\
\label{eq: thermo-diff-ratio-ion}
k_T^{(i)} =m_lm_h  \Bigl( \frac{c}{m_l} +  \frac{1-c}{m_h}    \Bigr) \Bigl[ \frac{cB_{ll}}{m_l} +\frac{(1-c)B_{lh}}{m_h}  \Bigr],\\
\label{eq: thermo-diff-ratio-el}
k_T^{(e)} = m_lm_h  \Bigl( \frac{c}{m_l} +  \frac{1-c}{m_h}    \Bigr) \Bigl[ \frac{cZ_l}{m_l} +\frac{(1-c)Z_h}{m_h}  \Bigr]  [(1-c)B_{le} - cB_{he} ] \frac{T_e}{T_i},
\end{eqnarray}
where quasi-neutrality condition was used to write Eq.~(\ref{eq: thermo-diff-ratio-el}). 

Expression~(\ref{eq: canonical-2}) does not have the exact form of
Eq.~(\ref{eq: canonical-flux-1}) since the $\nabla\log{T_e}$ term
appears on the right side. This is because the only external force
accounted for by Eq.~(\ref{eq: canonical-flux-1}) is the electric
field, whereas for the system of the two ion species considered here
the thermal force exerted by electrons is also external. Moreover,
equation~(\ref{eq: canonical-flux-1}) is only valid when at any given
point different components of the system are nearly equilibrated; in
particular, this means that temperatures of all the components must be
equal. For the system including ions only, this condition is satisfied
due to our ordering~(\ref{eq: ordering-time}). However, this ordering
does allow $T_e$ to be different from $T_i$, as the energy exchange
between the electron and any of the ion species takes longer than that
between the two ion species by a factor of
$\sqrt{m_{l,h}/m_e}$. Hence, even for the plasma as a whole, for which
the ion-electron thermal force is internal, $T_e$ and $T_i$ have to be set equal for thermodynamically obtained
Eq.~(\ref{eq: canonical-flux-1}) to be recovered. 
 It is interesting to note that $T_i\neq T_e$ is normally expected in an ICF capsule, especially
at the hot spot where fusion occurs.

Equations~(\ref{eq: diffusion-coeff})~and~(\ref{eq:
  electro-diff-ratio}) give the electro-diffusion coefficient $k_ED$,
thereby fulfilling the goal of this paper. Notice that $k_E$ goes to
zero if the charge-to-mass ratios are equal for the two ion
species. This result rigorously obtained here from ion fluid equations
has a simple physical explanation. Indeed, when $Z_l/m_l = Z_h/m_h$
the electric field does not distinguish between the light and heavy
ions and therefore does not contribute to the relative motion of the
species.

Unlike expressions~(\ref{eq: thermo-diff-ratio-ion})~-~(\ref{eq:
  thermo-diff-ratio-el}) for thermo-diffusion ratios, which involve
transport coefficients $B_{\alpha\beta}$, Eq.~(\ref{eq:
  electro-diff-ratio}) provides an explicit result for $k_E$ without
invoking a kinetic calculation. In other words, as its baro-diffusion
counterpart, the electro-diffusion ratio is a thermodynamic
quantity. Interestingly, it can then be evaluated in the same
relatively simple way as suggested in Ref.~\cite{zeldovich} for
evaluating the baro-diffusion ratio. We outline this calculation in
the appendix A.

\section{Discussion}

Relations~(\ref{eq: diffusion-coeff})~and~(\ref{eq:
  electro-diff-ratio}) do not provide an explicit result for the
electro-diffusion coefficient because of the transport coefficient
$A_{lh}$ entering the formula for $D$. However, as the approach presented
does provide an explicit result for $k_E$, a substantial insight into
the role of electro-diffusion can still be gained. 
To compare baro- and electro-diffusion caused perturbations of the
species concentrations we employ Eqs.~(\ref{eq:
  baro-diff-ratio})~and~(\ref{eq: electro-diff-ratio}) to evaluate the
ratio
\begin{equation}
\label{eq: electro-vs-baro}
\frac{k_E}{k_p}=\frac{Z_l/m_l - Z_h/m_h}{1/m_l-1/m_h} ,
\end{equation}
which depends on the properties of the ions only. In the special case
of the two isotopes of one element, i.e. $Z_l=Z_h \equiv Z$,
Eq.~(\ref{eq: electro-vs-baro}) gives $k_E/k_p=Z$. In particular, for
the practically important DT mix, the two coefficients turn out to be
equal, $k_E/k_p = 1.$ For the D\textsuperscript{3}He mix, commonly used to study
sub-ignited implosions, $k_E/k_p=-1$. In a plasma shock wave, the
electric field is directed towards the unshocked region to prevent
electrons' running ahead of ions and maintain quasi-neutrality. It can
therefore be observed that in the case of the DT mix baro- and
electro-diffusions act together, whereas in the case of the
D\textsuperscript{3}He mix the two tend to cancel each other.

Of course, relation between the baro- and electro-diffusion ratios
alone is not sufficient for relating the corresponding fluxes. The
total ion pressure gradient and the electric field also need to be
compared. While carrying out this comparison in a general case is
hardly possible, it is reasonable to assume $\nabla\log{p_i} \sim
e\nabla \Phi/T_i$. Moreover, in a shock wave, the
electric field is rather governed by the electron pressure
gradient. The pressure of electrons is often greater than that of ions
and therefore it is likely that the electro-diffusive flux may be
noticeably larger than the baro-diffusive flux. This becomes particularly
intriguing for the D\textsuperscript{3}He mix,  in which electro-diffusion
counteracts baro-diffusion. As a result,  \textsuperscript{3}He concentration may be increased over its unperturbed value, contrasting the neutral theory based expectation that it is the lighter fraction whose concentration is enhanced in the shock front~\cite{sherman}.

In terms of numerical modeling of the diffusive separation of the fuel ions
in ICF capsules, the most direct approach would be to solve the multi-component
plasma equations in its individual species form. The electric field is then explicitly evolved.
Alternatively, the ion fluid equations can be solved in the center of mass frame, i.e.
$\rho, {\bf u}, p_i,$ with the ion species concentration $c$ followed by Eq.~(\ref{eq:dcdt})
and Eq.~(\ref{eq: canonical-2}). With this approach, the electric field can be either independently evolved
using Maxwell's equations, or inferred from the equation of motion for the electrons in the quasineutral regime.
In this latter case, the electron inertia and electron viscosity are ignored, so
\begin{equation}
\label{eq:electric-field}
e \nabla\Phi = \frac{\nabla p_e}{n_e} - (B_{le} + B_{he}) \nabla T_e.
\end{equation} 
The above equation implies that, at a minimum, the fluid equations should evolve the electron temperature
separately from the ions', which fortunately is frequently done in ICF codes.

Finally, we comment on whether or not electro-diffusion, described
here by considering the ion species separately, can be attributed to
baro-diffusion in the plasma as a whole. As previously mentioned, the
plasma mass flux is essentially equal to the ion mass flux, because
the electron inertia is negligible. The question to be answered is
therefore whether or not the right side of Eq.~(\ref{eq: canonical-2})
can be represented in terms of the total plasma pressure gradient,
rather than in terms of the partial ion pressure gradient and the
electric field. To investigate the issue, we insert
Eq.~(\ref{eq:electric-field}) into Eq.~(\ref{eq: canonical-2}) to
obtain
\begin{equation}
\label{eq: total-baro-1}
{\bf i}_l = -\rho D\Bigl[\nabla c + \frac{k_p }{(n_l+n_h) T_i} \nabla p_i + \frac{k_E }{n_e T_i} \nabla p_e  
+ k_T^{(i)}\nabla\log T_i + \tilde{k_T}^{(e)} \nabla\log T_e \Bigr],
\end{equation}
where 
\begin{equation}
\tilde{k_T}^{(e)} = k_T^{(e)} - (B_{le} + B_{he}) (T_e/T_i)
k_E.
\end{equation} 
Eliminating both the electron and ion partial pressure gradients
in Eq.~(\ref{eq: total-baro-1}) by substituting the total pressure
gradient is only possible if
\begin{equation}
\label{eq: total-baro-2}
\frac{k_E}{k_p} = \frac{n_e}{n_l+n_h}
\end{equation}
for any values of $n_l$ and $n_h$. 

Combining Eq.~(\ref{eq: electro-vs-baro}) and the quasi-neutrality condition
$n_e = Z_ln_l + Z_hn_h,$ one finds that Eq.~(\ref{eq: total-baro-2}) may
be identically satisfied only for $Z_l = Z_h \equiv Z$. The right side
of Eq.~(\ref{eq: total-baro-2}) is then equal to $Z$ and indeed
matches the left side of Eq.~(\ref{eq: total-baro-2}) according to
Eq.~(\ref{eq: electro-vs-baro}). Employing $n_e/(n_l+n_h) = k_E/k_p =
Z $ in Eq.~(\ref{eq: total-baro-1}) we find
\begin{equation}
\label{eq: total-baro-3}
{\bf i}_l = -\rho D\Bigl[\nabla c + \tilde{k_p}\nabla\log p  
+ k_T^{(i)}\nabla\log T_i + \tilde{k_T}^{(e)} \nabla\log T_e \Bigr],
\end{equation}
where $p = p_i+p_e$ is the total plasma pressure and
\begin{equation}
\label{eq: baro-eff}
 \tilde{k_p} = (1+ZT_e/T_i)k_p.
\end{equation}
Equation~(\ref{eq: electro-vs-baro}) predicts a larger baro-diffusion coefficient, as compared to the case of a neutral binary mix. In the limiting case of $T_e=T_i$
$$
\tilde{k_p}(T_e=T_i) = (1 + Z) k_p,
$$
giving that the enhancement factor due to electro-diffusion is $(1+Z)$. This factor is familiar
from the well-known ambipolar enhancement for ion
diffusion with respect to the laboratory frame. However, here the impact
is on relative diffusion of two distinct ion species. 

In summary, representing the electric field effect on
inter-ion-species diffusion as a modification to the conventional
baro-diffusion coefficient is only possible when these ion species are in the
same charge state ($Z_l=Z_h$). Moreover, even in such a case, the electron
and ion temperatures need to be evolved separately for this effect to
be properly accounted for.

\begin{acknowledgements}

The authors wish to thank Peter Amendt of LLNL and Bhuvana Srinivasan of LANL for fruitful discussions,
Brian Albright of LANL for sharing an unpublished report on mass transport near high-Z/low-Z interfaces in
plasma media, and Russel Kulsrud of Princeton University for pointing out the relevance to stellar structure.

This work was supported by the Laboratory Directed Research and Development (LDRD) program of LANL.

\end{acknowledgements}

\appendix
\section{Evaluating electro-diffusion~ratio~in~the ~Zel'dovich-Raizer ~fashion}

In Ref.~\cite{landau-hydro} expression~(\ref{eq: baro-diff-ratio}) for
the baro-diffusion  ratio is obtained  by utilizing a  general formula
giving $k_p$  in terms of  the specific volume and  chemical potential
derivatives    over    concentration.    Instead,    Zel'dovich    and
Raizer~\cite{zeldovich}  notice  that once  $k_p$  is  known  to be  a
thermodynamic quantity,  the answer found in some  special case should
also work for all other  cases. In particular, evaluating $k_p$ can be
simplified by considering a  globally equilibrated system. Indeed, the
flux, as well  as the temperature gradient, is then  equal to zero and
the baro-diffusion ratio  can be obtained by balancing  the $\nabla c$
and  $\nabla\log{p}$   terms  on  the  right   side  of  Eq.~(\ref{eq:
  canonical-flux}). By writing  explicit expressions for the densities
of the mix components in the uniform gravitational field Eq.~(\ref{eq:
  baro-diff-ratio}) can  then be recovered. Below, we  apply this idea
to a plasma with two sorts of ions.

First, we recall Eq.~(\ref{eq: canonical-flux-1}) and set $\nabla T_i=0={\bf i}_{l,h}$  to obtain
\begin{equation}
\label{eq: diffusion-equilibrium}
 \frac{dc}{dx} +k_p \frac{ d\log{p_i}}{dx} + \frac{e k_E}{T_i}\frac{d\Phi}{dx} =0,
\end{equation}
where $k_p$ is readily provided by Eq.~(\ref{eq: baro-diff-ratio}). Next, we notice that for a plasma equilibrated in the uniform gravitational field, the light and heavy ion density profiles are given by
\begin{equation}
\label{eq: barometric}
n_{l,h} =n_{l0,h0}\exp{\Bigl(-\frac{m_{l,h}gx}{T_i}-\frac{Z_{l,h}e\Phi}{T_i}\Bigr)},
\end{equation}
where $n_{l0,h0}$ are the species number densities at $x=0$ and $\Phi(x=0)=0$. Now, the first two terms on the left side of Eq.~(\ref{eq: diffusion-equilibrium}) need to be evaluated with the help of Eq.~(\ref{eq: barometric}).

To do so, we observe that $c = m_ln_l/(m_ln_l+m_hn_h)$ to write
\begin{equation}
\label{eq: grad-f}
 \frac{dc}{dx} = -c^2\frac{d}{dx}\Bigl( \frac{1}{c}  \Bigr) = -c^2\frac{d}{dx}\Bigl( \frac{m_hn_h}{m_ln_l}  \Bigr).
\end{equation}
Next, we insert Eq.~(\ref{eq: barometric}) into the right side of Eq.~(\ref{eq: grad-f}) to find
\begin{equation}
\label{eq: grad-f-1}
 \frac{dc}{dx} = -c(1-c) \Bigl[ -\frac{g}{T_i}(m_h-m_l) +  \frac{eE}{T_i}(Z_h-Z_l)       \Bigl].
\end{equation}
Finally, the total ion pressure gradient is calculated along the same lines to obtain
\begin{equation}
\label{eq: grad-p}
\frac{ d\log{p_i}}{dx} = -\frac{g}{T_i}\frac{(m_ln_l+m_hn_h)}{n_l+n_h} + \frac{eE}{T_i}\frac{(Z_ln_l+Z_hn_h)}{n_l+n_h}.
\end{equation}
Then, by inserting Eqs.~(\ref{eq: baro-diff-ratio}),~(\ref{eq: grad-f-1}) and~(\ref{eq: grad-p}) into Eq.~(\ref{eq: diffusion-equilibrium}) and solving it for $k_E$, previously obtained result~(\ref{eq: electro-diff-ratio}) is reproduced.


\begin{thebibliography}{10}
\bibitem{note-yield} If instead the laser driver rather than the target temperature is held fixed,
optimal fusion yield can be obtained with different reactant number densities
which leads to a higher target temperature upon implosion, espcially in sub-ignition plasmas.
\end{thebibliography}
\end{document}